\documentclass[pre,amsmath,amssymb,twocolumn,superscriptaddress]{revtex4-1}
\usepackage{amsmath,amssymb,graphicx,bm}

\newcommand{\xiamen}{Department of Physics, Key Laboratory of Low Dimensional
Condensed Matter Physics (Department of Education of Fujian Province), and
Jiujiang Research Institute, Xiamen University, Xiamen 361005, Fujian, China}
\newcommand{\fuzhou}{School of Science, Jimei University, Xiamen 361021, China
and Department of Physics, Fuzhou University, Fuzhou 350108, Fujian,
China}

\begin{document}

\title{Thermal rectification in the thermodynamic limit}
\author{Shunjiang You}
\affiliation{\xiamen}
\author{Daxing Xiong}
\email{phyxiongdx@fzu.edu.cn; xmuxdx@163.com}
\affiliation{\fuzhou}
\author{Jiao Wang}
\email{phywangj@xmu.edu.cn}
\affiliation{\xiamen}

\begin{abstract}
We show that an increasingly strong thermal rectification effect occurs
in the thermodynamic limit in a one-dimensional, graded rotor lattice with
nearest-neighboring interactions only. The underlying mechanism is related
to the transition from normal to abnormal heat conduction behavior observed
in the corresponding homogeneous lattices as the temperature decreases. In
contrast and in addition to that by invoking long-range interactions, this
finding provides a distinct scenario to make the thermal rectification
effect robust.
\end{abstract}

\maketitle

Thermal rectification (TR) refers to the effect whereby heat current
flows across a system preferably in a certain direction rather than in the opposite direction. As the thermal counterpart of the rectification of electric current, it can be envisaged that TR may play a vital role in thermal control
and management. TR is also very interesting from the viewpoint of the
fundamental theories. In fact, the study of this effect has a long history
(see, e.g., Ref.~\cite{Walker} for a review) and recently, it has aroused
recurrent interest since the seminal work by Terraneo {\it et al.}, who
tried to reveal the underlying mechanism based on the microscopic
dynamics~\cite{PeyrardPRL}. Encouraged by the experimental verification
of TR with nanotubes~\cite{CWChang}, it is believed that, eventually,
the efforts invested in studying this effect will result in novel
technologies and applications~\cite{Modrev}.

So far two general, necessary conditions for TR have been identified.
Above all, the system must have an asymmetric structure along the direction
of the heat current. In this respect, two typical asymmetric structures have
been put forward and investigated both intensively and extensively. One is
the hybrid structure of sequentially coupled two or three segments different
in nature~\cite{PeyrardPRL, PeyrardEPL, Rev2016, XXZ, Small}, while another
class features a mass- or geometry-graded configuration~\cite{CWChang, YNPRB,
EP2010, JW2012, JPA2013}. The second ingredient of TR is that the local heat transport behavior should depend on the local structure
and temperature~\cite{PeyrardEPL}. In this respect, the temperature dependence
of the heat conductivity, in particular for TR in a lattice system, has
been scrutinized carefully. In general, a sensitive temperature dependence
of the heat conductivity is favorable. Interesting examples were proposed
in Ref.~\cite{Kobayashi, Small},  where a critical phase-transition point
of the heat conductivity was exploited.

The mechanisms for fulfilling these two general conditions to obtain TR
are various~\cite{Liu, Eckmann, Casati, JPCB, JPCC}. However, in spite of
the progresses made, these mechanisms have not yet been satisfactorily
understood, particularly in systems such as insulating crystals where
the thermal energy is transported mainly through the lattice vibrations. In
this situation, an outstanding puzzle is why the TR effect is observed to
decay as the system size increases and eventually vanish in the thermodynamic
limit~\cite{YZh2006}. To prevent the decay, graded models with
long-range interactions were recently proposed based on linear-response
theory~\cite{EP2013}, and its effectiveness was verified numerically in
both the linear-response regime and beyond~\cite{SD2015}. These studies have
suggested the important role of long-range interactions for generating the
enhanced TR effect.

In this work, we reveal a distinct mechanism that may not only prevent the
decay of TR, but even make it progressively stronger in the thermodynamic
limit. We adopt the rotor lattice model~\cite{RL-Benettin, RL-Livi}, where
the component rotors can both vibrate and rotate with respect to their
equilibrium angular positions (see Fig.~1). We show that it is possible to
design a graded rotor lattice where TR increases as the system size, even
if the interactions act only on the nearest-neighboring rotors.

\begin{figure}[!b]
\includegraphics[width=8.0cm]{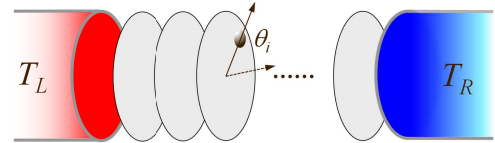}
\caption{A schematic plot of the rotor lattice that couples with two heat baths
at its two ends. For visualization purposes, a rotor is represented by a mass
point fixed on a rigid, massless disk that vibrates and rotates freely around
the horizontal axis. The angular variable $\theta_i$ of the $i$th rotor is
measured as the angle formed between the line connecting the center of the
disk and the mass point and that represents the given reference direction
(the dashed arrow line).}
\end{figure}

\begin{figure*}[!]
\vskip 0.5cm
\includegraphics[width=18.0cm]{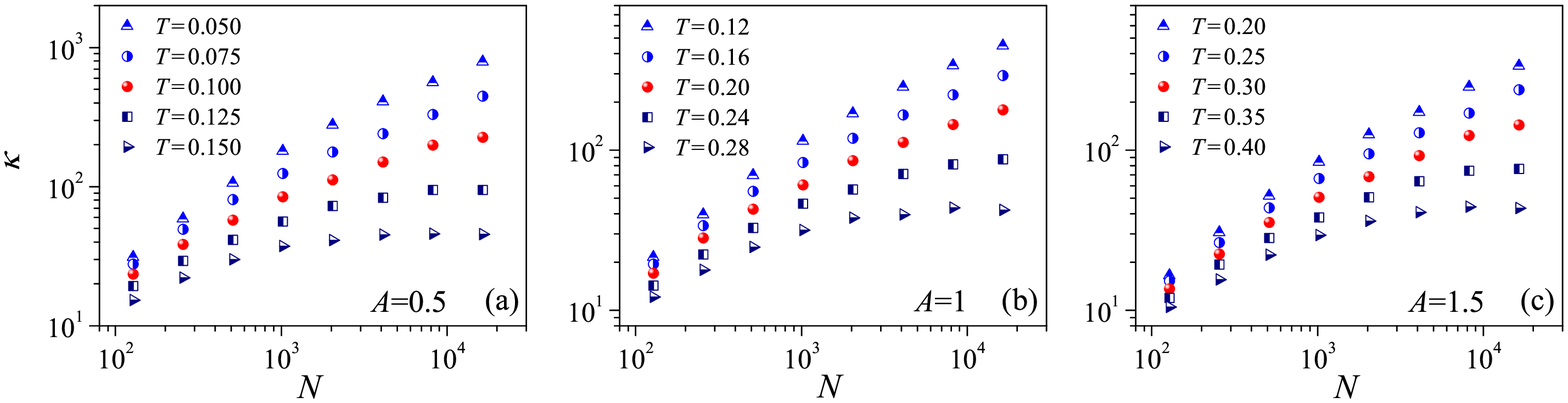}
\caption{The heat conductivity $\kappa$ of the homogeneous rotor lattice
system as a function of the system size $N$ at various temperatures for
$A$=0.5 (a), $A$=1 (b), and $A$=1.5 (c), respectively. It can be seen
that the transition temperature $T^{tr}\approx A/5$, below which the heat
conductivity tends to diverge and above which the heat conductivity tends
to saturate as the system size increases.}
\end{figure*}

In our model, the rotors are assumed to be positioned on a one-dimensional lattice with a unity lattice constant (we use dimensionless variables
throughout). For a homogeneous rotor lattice, its Hamiltonian is
\begin{equation}
H=\sum_i [\frac{1}{2}\dot{\theta}^2_i+V(\theta_{i+1}-\theta_i)],
\end{equation}
where $\theta_i$ is the angular variable of the $i$th rotor with respect
to a given reference axis, $\dot{\theta}_i$ is the corresponding conjugate
variable, and $V(x)$ is a periodic potential whose conventional version
is $V(x)=1-\cos(x)$. This model was put forward as a prototype for spin
systems of statistical mechanics~\cite{RL-Benettin} and was found relevant
in applications (e.g., its potential describes the relative rotation
of nearest-neighboring polymer fragments around the axis of a
macromolecule~\cite{RL-Gendelman} and interestingly, relevant to our
topic here, it has been shown recently that a strong TR effect is possible
in a polyethylene nanofiber system~\cite{Small}). A striking property of
the rotor lattice is that its heat conduction behavior may undergo a
transition~\cite{RL-Giardina, RL-Savin}: In the low-temperature regime,
the rotors only vibrate around their equilibrium positions and the heat
is transported dominantly by phonons. As a result, the heat conductivity
increases with system size, showing an abnormal heat conduction behavior.
But in the high-temperature regime, nonlinear modes due to rotations
emerge and play a role; consequently the heat conductivity will converge
to a constant as the system size increases, so that the system has a normal
heat conduction behavior as described by the Fourier law.

Our motivation is to take advantage of this transition for generating
enhanced TR. To this end, we first turn to a variant homogeneous
rotor lattice with potential
\begin{equation}
V(x)=A[1-\cos(\omega x)],
\end{equation}
where $A$ and $\omega$ are two parameters. To keep the maximum
force $|F_{max}|$ between any two neighboring rotors a constant, we set
$\omega=1/A$ such that $|F_{max}|=1$, so that $A$ serves as the only
independent parameter. First of all, we investigate how the transition temperature $T^{tr}$ depends on the parameter $A$~\cite{note}. The heat
conductivity is considered to characterize the system's heat conduction
behavior, which is obtained numerically by following the conventional
nonequilibrium simulation method. Namely, for a system of $N+1$ rotors
at a given temperature $T$, its two ending rotors, the zeroth and the $N$th,
are coupled with two Langevin thermostats of temperature $T_L=T+\Delta T$
and $T_R=T-\Delta T$, respectively, whose motion equations are
$\ddot{\theta}_0 =-\partial V(\theta_1-\theta_0)/\partial \theta_0-\gamma
\dot{\theta}_0+\xi_0$
and
$\ddot{\theta}_{N}=-\partial V(\theta_{N}-\theta_{N-1})/\partial \theta_{N}
-\gamma \dot{\theta}_{N}+\xi_{N}$,
with $\xi_0$ and $\xi_{N}$ being two white Gaussian noises satisfying
$\langle\xi_0(t)\xi_0(t') \rangle=2\gamma k_B T_L \delta(t-t')$
and
$\langle\xi_{N}(t)\xi_{N}(t')\rangle= 2\gamma k_B T_R \delta(t-t')$.
The $N-1$ rotors in between follow their own motion equations determined
by the Hamiltonian only. The system is evolved numerically
from an initial condition given arbitrarily until the stationary state
is reached, then the time-averaged heat current is measured according
to the definition $j=\langle j_i\rangle=\langle-[\partial V(\theta_{i+1}
-\theta_i)/\partial\theta_i +\partial V(\theta_{i}-\theta_{i-1})/\partial
\theta_i]\dot{\theta}_i/2\rangle$.
The heat conductivity $\kappa$ is thus obtained by assuming the Fourier
law $j=-\kappa \nabla T$, which leads to $\kappa \approx jN/(T_L-T_R)
=jN/\Delta T$. In our simulations the Boltzmann constant $k_B$ is set
to be unity throughout and, when the heat baths are involved, the coupling
constant is set to be $\gamma=1$. We have also verified that when more
boundary rotors are thermalized with the baths and when different values
of $\gamma$ are taken, the results remain the same.

The numerical results of the heat conductivity are presented in Fig.~2. We
find that the transition temperature depends linearly on the parameter $A$ ,
i.e., $T^{tr}\approx A/5$, which can be related to the temperature value
at which the force between two neighboring rotors may reach its maximum
value allowed (see below). In our simulations for Fig.~2, the temperature
bias is set to be $\Delta T=T/5$, which has been verified to be small enough
to ensure that the system falls in the linear-response regime. Note that,
in Fig.~2 and all other figures, the errors of the data are less than $1\%$
and the error bars are smaller than the data symbols, hence the error
bars are omitted throughout.

At a low temperature, the rotor lattice reduces to the Fermi-Pasta-Ulum
lattice~\cite{Spohn, Dhar} and the heat conductivity diverges in the
thermodynamic limit~\cite{FPU}. As the temperature increases to exceed $T^{tr}$,
localized nonlinear rotational modes~\cite{breather} form and interact with
the heat current to prevent its free propagation~\cite{RL-Savin}, bringing
the system into the normal diffusive transport regime~\cite{Dhar}. Hence as
an estimation of $T^{tr}$, we can take advantage of the necessary condition
that a rotor begins to rotate around its neighbors when the temperature is
increased, which leads to that at $T^{tr}$, the relative angular displacement
between two neighboring rotors satisfies $\omega (\theta_{i+1}-\theta_{i})
=\pi/2$, or, equivalently, the force between them reaches $|F_{max}|$. Empirically,
we can assume that for $T<T^{tr}$ the force between two neighboring rotors
follows the Gaussian distribution and, at $T=T^{tr}$, the width of the
distribution reaches $|F_{max}|$. Then, by identifying $|F_{max}|$ with the
full width at half maximum of the force distribution, i.e., $2\sqrt{2\ln2}
\sigma_F$ (with $\sigma_F$ being the standard deviation)~\cite{FWHM}, we find
that $T^{tr}\approx A/5$ over a wide range of $A$ ($0.1<A<10$) that has
been investigated, which agrees very well with the value of $T^{tr}$ obtained
based on the $\kappa$ versus $N$ relations (see Fig.~2). As an example, the
force distribution for $A=1$ is shown in Fig.~3.

\begin{figure}[!t]
\includegraphics[width=8.0cm]{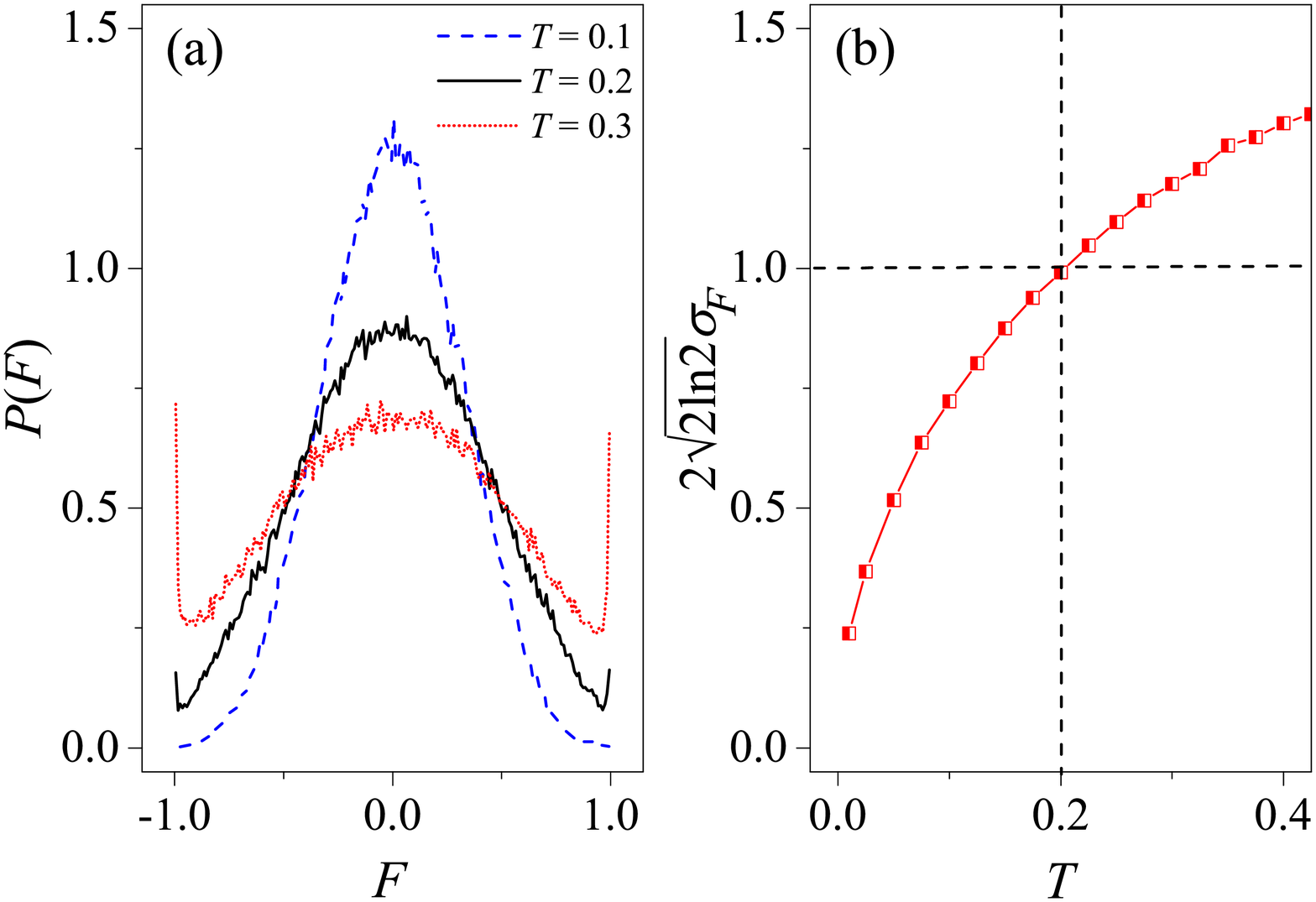}
\caption{(a) The distribution of the force acts on a rotor in the homogeneous
rotor lattice with $A=1$ when the system is at the equilibrium state of
temperature $T$. (b) The full width at half maximum of the force distribution
as a function of the temperature. The transition temperature is estimated
empirically by identifying it with $|F_{max}|=1$, which results in $T^{tr}
\approx 0.2$ in this case for $A=1$.}
\end{figure}

Based on the relation between $T^{tr}$ and $A$, we can design a graded rotor
lattice system with graded potentials, of which that between the $i$th and
the$(i+1)$st rotor is
\begin{equation}
V_i=A_i\{1-\cos[\omega_i(\theta_{i+1}-\theta_{i})]\},
\end{equation}
where $\omega_i$ is fixed to be $\omega_i=1/A_i$ and $A_i$ changes linearly
from $A_0=A_L$ to $A_N=A_R$. Denote the transition temperature in the homogeneous
system of $A=A_\alpha$ by $T^{tr}_{\alpha}$ (with $\alpha=L$ and $R$) and, without loss of generality, let us suppose $A_L>A_R$, so that $T^{tr}_{L}>T^{tr}_{R}$. Because the system has an asymmetric structure and the heat conduction depends on the local temperature and interactions, the TR effect is expected.

Suppose the ``working'' temperature range of the system is $[T_{-}, T_{+}]$,
we can measure the rectification factor~\cite{Walker}
\begin{equation}
f=\frac{|\kappa_f-\kappa_r|}{\kappa_f+\kappa_r},
\end{equation}
where $k_f$ is the forward heat conductivity measured when the temperatures
of the heat baths are set to be $T_L = T_{+}$ and $T_R = T_{-}$  and $\kappa_r$
is the reverse one measured with the swopped heat baths, i.e., $T_L = T_{-}$
and $T_R = T_{+}$. This definition is equivalent to $f={|j_f-j_r|}/(j_f+j_r)$
with $j_f$ and $j_r$ being the corresponding forward and backward heat current.

In the limit that interests us of large system size, a lattice can be treated
as a number of sequential segments. It is partitioned in such a way that each
segment is sufficiently small to justify the local equilibrium assumption,
allowing the temperature and interaction gradient over it to be neglected and
each segment to assume a certain local temperature and $A$ value. On the other
hand, the segments are partitioned large enough so that each one can be assigned
a local transition temperature, around which the segment's heat conduction may
change drastically. It is worth noting that, the larger the system, the larger
the segments, then the more abrupt this change (see Fig.~2). Hence an increasingly enhanced TR effect should be possible in the thermodynamic limit. This is the most intriguing property to be expected.

\begin{figure}[!t]
\includegraphics[width=8.0cm]{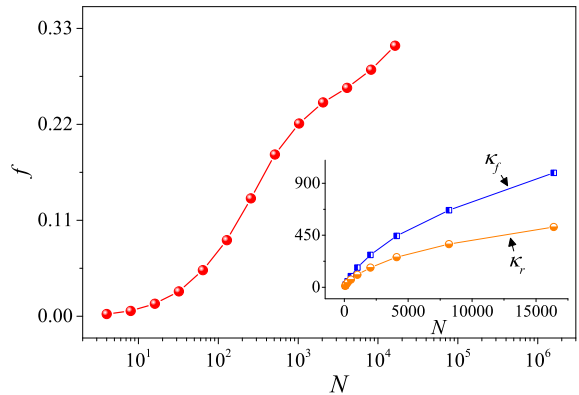}
\caption{The rectification factor as a function of the system size in the
graded rotor lattice for the case that $T^{tr}_{L}>T_{+}>T^{tr}_{R}>T_{-}$.
Here $T_{+}=0.25$, $T_{-}=0.05$, $A_{L}=1.5$ ($T_L^{tr}\approx 0.3$ ), and
$A_{R}=0.5$ ($T_R^{tr}\approx 0.1$). The inset shows the heat conductivities
evaluated with the forward and the reversed heat current, respectively.}
\end{figure}

Nevertheless, besides the system size, TR in our system also depends
on the working temperature range. For example, for the case
$T^{tr}_{L}>T^{tr}_{R}>T_{+}>T_{-}$, the local temperature of any segment is
below its transition temperature so that both $\kappa_f$ and $\kappa_r$
are large. As a consequence, $f$ should be low. Similarly, for
$T_{+}>T_{-}>T^{L}_{tr}>T^{R}_{tr}$, the local temperature of any segment is
above its transition temperature, so both $\kappa_f$ and $\kappa_r$
are small and $f$ should be low as well. The most interesting case is for
$T^{tr}_{L}>T_{+}>T^{tr}_{R}>T_{-}$, where for the forward current, because
$T^{tr}_{L}>T_{L}>T^{tr}_{R}>T_{R}$, every segment has a temperature lower than
its transition temperature, so that $\kappa_f$ is large. But for the reversed
current, because $T_{R}=T_{+}>T^{tr}_{R}$, the segment(s) at the right end of the
system is (are) above its transition temperature, so $\kappa_r$ is small.
As a result, the rectification effect should be remarkable. For other working
temperature ranges, the TR factor should be in between these two limits.

The simulation results corroborate these conjectures very well. First of
all, in Fig.~4 is shown the rectification as a function of system size for the
optimal case $T^{tr}_{L}>T_{+}>T^{tr}_{R}>T_{-}$. It can be
seen that, indeed, the rectification increases with system size and reaches
the considerably large value of $f\approx 0.3$ at the system size $N\sim 10^4$.
In view of the fact that $A_{L}$ and $A_{R}$ are fixed so the interaction
gradient actually decays linearly with the system size, TR in our model
is particularly striking compared with that observed with the fixed mass
gradient in oscillator lattices of long-range interactions~\cite{SD2015}.
The rectification shown in Fig.~4 is expected to increase further for
$N> 10^4$, but unfortunately, the simulations have turned out to be
prohibitively expensive. In spite of this, the robust increasing trend
with a roughly constant rate can be clearly recognized from the available
data. In the inset of Fig.~4 are shown the forward and the reverse heat conductivity. It can be seen that the latter has not saturated yet at the
largest system size $N\approx 1.6 \times 10^4$ simulated. However, to
increase $N$ further, it is expected that $\kappa_r$ will saturate while
$\kappa_f$ keeps increasing to diverge.

\begin{figure}[!t]
\includegraphics[width=8.0cm]{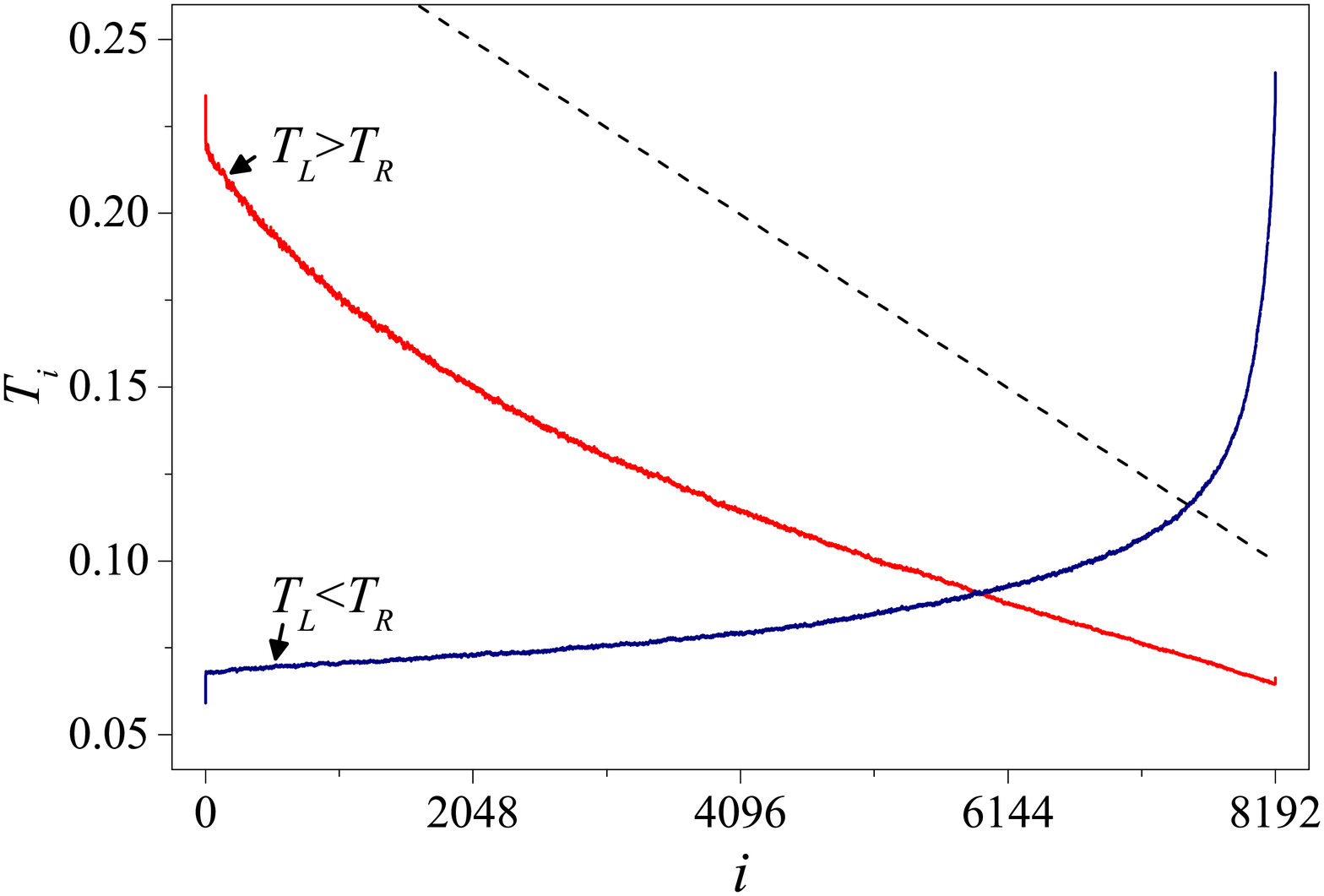}
\caption{The temperature profile of the graded rotor lattice for $T_{-}=0.05$
and $T_{+}=0.25$ when the stationary heat current is formed with $T_L=T_{+}>
T_R=T_{-}$ (forward current, red curve) and reversely with $T_L=T_{-}< T_R=
T_{+}$ (reverse current, dark blue curve). The dashed line indicate the
transition temperature in a homogeneous lattice with $A=A_i$. Here $A_L=1.5$,
$A_R=0.5$, and the system size is $N=8192$.}
\end{figure}

The temperature profiles provide more details of the mechanism of TR in our
model. As an example, those corresponding to the data point next to the last
one in Fig.~4 are plotted in Fig.~5. As shown, for the forward current,
the local temperature is lower than the transition temperature all through
the lattice, which is consistent with the strong forward current. In
contrast, for
the reverse current, the temperature changes mainly over the last
segment, suggesting that its contribution to the thermal resistance is dominant.
Therefore, it is the last segment that dominantly restricts the reverse
heat current. As the system size increases, this segment grows accordingly;
hence its local conductivity, and thus $\kappa_r$ for the whole system,
would saturate eventually.

\begin{figure}[!t]
\includegraphics[width=8.0cm]{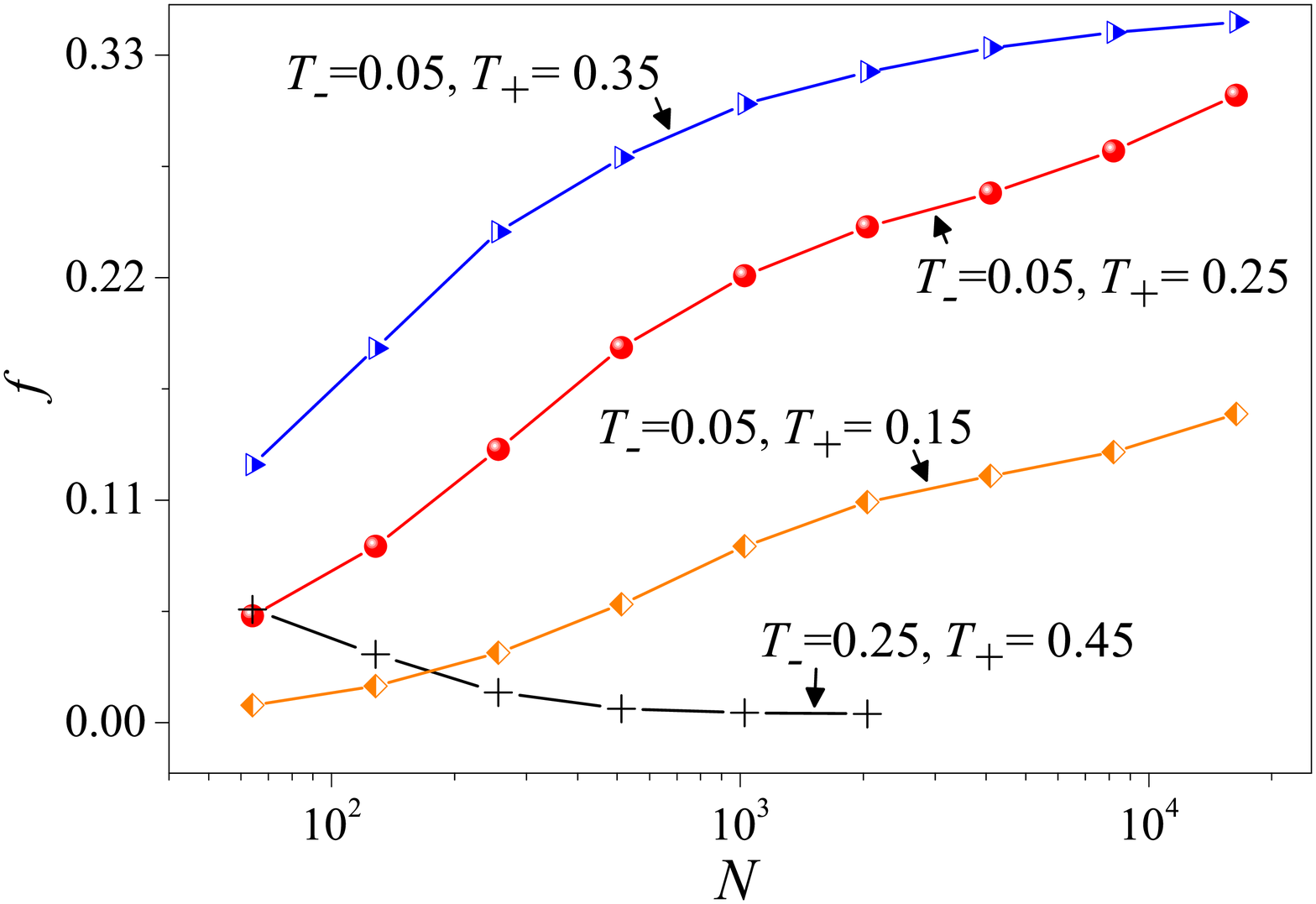}
\caption{The rectification factor as a function of the system size in the
graded rotor lattice in various working temperature ranges. Here $A_L=1.5$
and $A_R=0.5$.}
\end{figure}

A thorough comparison of different working temperature ranges is made in Fig.~6.
That shown in Fig.~4 for $T_{-}=0.05$ and $T_{+}=0.25$ is reproduced here as well.
Besides it, that with $T_{-}=0.05$ and $T_{+}=0.15$ also belongs to the optimal
case, but with a smaller temperature bias ($T_{+}-T_{-}=0.1$). In this case,
as expected, the rectification at a given $N$ is smaller, but the rectification
should increase further with system size. The case of $T_{-}=0.05$ and
$T_{+}=0.35$ corresponds to the largest temperature bias, which is why, at small $N$, its rectification is the strongest. However, because
$T_{+}>T^{tr}_{L}$, for the forward current, the left-most segment works at
a local temperature higher than $T^{tr}_{L}$, its conductivity will saturate
in the limit of large system size. Indeed, the simulation results suggest that the increasing rate decreases progressively with increasing $N$ in the investigated range of $N$. For the last example case of $T_{-}=0.25$ and $T_{+}=0.45$, the whole lattice is above the local transition temperature,
so both the forward and reverse current are weak, which explains the decaying rectification.

Finally, note that, in the studies of Fig.~6, the temperature biases adopted
are strong in the sense that $T_{+}-T_{-}$ is comparable with the nominal
temperature $T=(T_{+}+T_{-})/2$ of the system. As expected, for a fixed
system size, the rectification factor in general decreases as the the temperature bias decreases, which is also the case in our model. However, as long as the condition
$T^{tr}_{L}>T_{+}>T^{tr}_{R}>T_{-}$ is respected, a divergent rectification factor in the thermodynamic limit can still be predict, just as shown by the case of $T_{-}=0.05$ and $T_{+}=0.15$. It suggests that a strong TR effect is in principle possible even when the temperature bias tends to zero, i.e., $T_{+}$ and $T_{-}$ approach $T^{tr}_{R}$ from above and below, respectively, given
a large system size.

In summary, we have shown that, in the interaction-graded rotor lattice
with only nearest-neighboring interactions, a progressively enhanced TR
effect is possible in the thermodynamic limit. Unlike the previously proposed
mechanisms, such as matching and dismatching of phonon bands in
sequentially assembled lattices~\cite{PeyrardPRL} and that of long-range
interaction-induced transport channel asymmetry~\cite{EP2013,SD2015}, in
our model TR results from the suppression of heat conduction by
the local nonlinear modes stimulated at one end of the system when being
coupled to the high temperature bath. Here the local nonlinear modes are
due to the relative rotation motion between the rotors; but it is
interesting to note that structural rotations -- e.g., that between segments
of polyethylene nanofibers as a response to increasing temperature causing
a morphology change from an all-trans conformation crystalline structure
to a disordered structure~\cite{Small} -- may lead to sensitive temperature-dependent heat conduction and can be utilized to generate TR as
well~\cite{Small}. It suggests that other nonlinear modes, which may enhance
or suppress heat conduction and meanwhile have a sensitive temperature
dependence, deserve further investigations. It is also interesting to
study if introducing long-range interactions may further improve
TR. In this respect, it raises the question of how long-range interactions
may affect the nonlinear modes and their implications on heat conduction,
and the rotor lattice may serve as an ideal paradigmatic model again.

We acknowledge support by the NSFC (Grants No. 11535011, No. 11335006, and
No. 11575046), the NSF of Fujian Province of China (Grant No. 2017J06002),
and the Qishan scholar research fund of Fuzhou University.




\begin{thebibliography}{100}
\bibitem{note}
Note that the transition in the rotor lattice may not be a phase transition.
Therefore $T^{tr}$ should be understood as a representative value of the
temperature range over which the heat conduction behavior undergoes the
qualitative change as shown in Fig.~2.
\end{thebibliography}
\end{document}